\documentclass{nature_wFig}
\usepackage{graphicx}
\title{Chiral optical response of planar and symmetric nano\-trimers enabled by heteromaterial selection}

\author{Peter Banzer$^{1,2,3,4,*}$, Pawel Wozniak$^{1,2,4,*}$, Uwe Mick$^{1,2,4}$, Israel De Leon$^{3,4}$ \& Robert W. Boyd$^{3,4,5}$}

\begin{document}

\maketitle
\begin{affiliations}
\item Max Planck Institute for the Science of Light, Guenther-Scharowsky-Str. 1, D-91058 Erlangen, Germany
\item Institute of Optics, Information and Photonics, University Erlangen-Nuremberg, Staudtstr. 7/B2, D-91058 Erlangen, Germany
\item Department of Physics, University of Ottawa, 25 Templeton, Ottawa, Ontario K1N 6N5 Canada
\item Max Planck University of Ottawa Centre for Extreme and Quantum Photonics, University of Ottawa, 25 Templeton Street, Ottawa, Ontario, K1N 6N5 Canada
\item Institute of Optics, University of Rochester, Rochester, New York, 14627, USA
\end{affiliations}

\begin{abstract}
Chirality is an intriguing property of certain molecules, materials or artificial nanostructures, which allows them to interact with the spin angular momentum of the impinging light field. Due to their chiral geometry, they can distinguish between left- and right-hand circular polarization states or convert them into each other. Here, we introduce a novel approach towards optical chirality, which is observed in individual two-dimensional and geometrically mirror-symmetric nanostructures. In this scheme, the chiral optical response is induced by the chosen heterogeneous material composition of a particle assembly and the corresponding resonance behavior of the constituents it is built from, which breaks the symmetry of the system. As a proof of principle, we investigate such a structure composed of individual silicon and gold nanoparticles both experimentally as well as numerically. Our proposed concept constitutes a novel approach for designing two-dimensional chiral media tailored at the nanoscale.
\end{abstract}

Optical chirality is a well-known phenomenon that has been intensely studied in the past and still attracts tremendous attention, especially in the field of modern optics. Optically chiral media possess a sense of handedness that allows them to interact differently with left- and right-handed circularly polarized light, and gives rise to important optical phenomena such as polarization rotation and circular dichroism (CD). Natural chiral media are composed of chiral molecules. The chiral optical response of many chiral molecules is understood as a direct consequence of their three-dimensional chiral geometry, which does not allow for a superposition of the molecule's chiral structure with its mirror-image. Nowadays, modern micro- and nano-fabrication techniques allow for the creation of elaborately and artistically shaped structures, such as three-dimensional helical antennas\cite{Gansel2009,Thiel2010,Hoeflich2011,Helgert2011,Frank2013,Belardini2014,Esposito2015}. Such structures exhibit a strong chiral response, especially when arranged into dense lattices. In particular, microscopic or nanoscopic helices are a prototypical example of artificial chiral structures, following one of mother nature's design recipes for chiral molecules. Their chiral optical response can be tuned or artificially created by controlling their size, the number of pitches or twists around the central axis and other geometric parameters. In contrast to their molecular archetypes, which are orders of magnitudes smaller than the wavelength of light, artificial chiral structures are comparable in size with the wavelength. As a consequence, the instantaneous electric field points into different transverse directions at different cross-sections along the helical object. Hence, such artificial structures may exhibit a richer resonance behavior, which manifests itself in a large number of supported resonant (higher-order) modes and a stronger chiral response. Reviews on optical chirality and optically chiral nanostructures have been presented recently\cite{Schaeferling2012,Valev2013}. Furthermore, this introduction wouldn't be complete without mentioning important theoretical contributions describing optical chirality in general, and the interaction of light's spin angular momentum with matter in particular\cite{Tang2010,Bliokh2011,Coles2012,Fernandez2013}.\\

\begin{figure}
\includegraphics[width=0.75\textwidth]{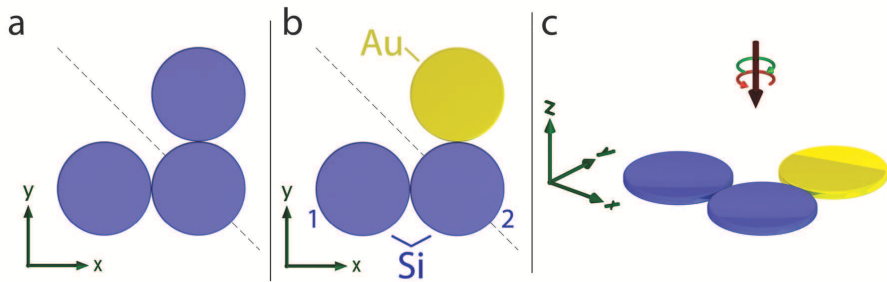}
\caption{\label{fig:Fig1}\textbf{Illustration of the investigated heterogeneously composed nanodisk trimer.} (a) Sketch of a three-particle system (trimer) consisting of nanodisks of the same size and material all placed in the same plane. (b) If one of the particles at the end of the trimer is replaced by an equally sized disk made from a different material, the trimer still exhibits geometrical mirror-symmetry (gray dashed line indicates its geometric symmetry axis), but with respect to the material composition, its symmetry is broken. The trimeric nanostructure consists of three nanodisks, two made of silicon (blue) and one made of gold (yellow). (c) In the simulations, the trimer is illuminated with circularly polarized light (Gaussian beam), which impinges normally (along the negative $z$-direction) to the plane defined by the three disks ($xy$-plane). The thickness (height) and diameter of the disks were chosen to be 30 nm and 180 nm, respectively.}
\end{figure}

It also has been shown that a similar chiral optical behavior can result from a three-dimensional (3D) chiral or layered arrangement of achiral nanoparticles, such as nanospheres, split-ring resonators or similar\cite{Liu2007,Liu2009,Decker2009,Fan2010,Hoeflich2011,Kuzyk2012,Schaeferling2012,Hentschel2012,Shen2013,Kaell2014}. To showcase a comprehensible example here, one might think about an ensemble of nanospheres following a helical trace in 3D-space\cite{Fan2010,Hoeflich2011,Kuzyk2012,Schaeferling2012,Hentschel2012,Shen2013}. The resulting particle ensembles exhibit an intrinsically chiral geometry when effectively considered as a single object.\\ Truly planar or two-dimensional structures can also show a chiral optical response. Similar to the aforementioned 3D-chiral approaches, planar structures obtain their chiral optical properties by their lack of in-plane mirror-symmetry. Prominent examples are $S-$ and $L-$shaped antennas or gammadion-type nanostructures\cite{Papakostas2003,Rogacheva2006,Reichelt2006,Decker2007,Plum2009a,Hendry2010,Zhao2011,Eftekhari2012,Narushima2014,Alali2014}. Many such examples have been discussed in literature. Nonetheless, the full capabilities and underlying processes of chiral light-matter interaction, especially for planar chiral nanostructures, haven't yet been fully explored.\\
To obtain a chiral optical response from a two-dimensional nanostructure, also an alternative route has been proposed, which is based on so-called extrinsic or pseudo-chirality\cite{Plum2008,Plum2009,Plum2009a,Sersic2011,DeLeon2015}. In contrast to the intrinsically geometrically chiral structures discussed so far, the concept of extrinsic chirality allows for a chiral optical response of symmetric and achiral nanostructures. In this scheme, an array of achiral nanostructures, such as split-ring resonators, is illuminated under oblique incidence. In this situation, the structures are not super-imposable with their mirror-images anymore with respect to a rotation restricted to a plane perpendicular to the propagation direction of the impinging light, rendering them chiral.\\
It is worth noting here that the majority of the above-mentioned systems were comprised of lattices (metasurfaces) of individual nanostructures or were investigated in an inhomogeneous environment (on substrates). Both settings are known for being capable of enhancing the chiral optical effect. For instance, the presence of a substrate breaks the symmetry of the system along the surface normal, hence enhancing the chiral effect by making the lower and upper surface of a two-dimensional structure distinguishable.\\
Yet another approach towards optical chirality, which is of particular importance for this Letter, is the use of heterogeneous material compositions, similar to chiral molecules, which are built from different atoms. Corresponding examples reported in literature to date can be sub-categorized into two different geometries. First, those approaches based on structures exhibiting a three-dimensional heterogeneous material composition, which gain their chiral optical response dominantly from their three-dimensional chiral geometry rather than by their heterogeneity alone\cite{Yeom2013,Hentschel2013,Kaell2014}. And second, a theoretical approach based on quasi-2D nanostructures exhibiting an inhomogeneous and asymmetric material composition, while also lacking symmetry with respect to their shape and geometry\cite{Alali2014}. Both categories have been studied for arrays forming lattices or metasurfaces of the corresponding structures only.\\
In this Letter, we propose a truly planar nanostructure, which is geometrically achiral (mirror-symmetric) but exhibits a significant chiral optical response, i.e. circular dichroism, due to its heterogeneous composition. This response is observed for an individual nanostructure (as opposed to the arrangement in an array or a metasurface), which is embedded homogeneously into a dielectric medium and illuminated under normal incidence. This is in striking contrast to other two-dimensional chiral structures discussed in literature so far. In the presented case, the structural symmetry is broken, and a handedness or chiral response is induced, by the heterogeneous material composition of the assembly (see Fig. \ref{fig:Fig1}). With respect to its geometry, the nanostructure, assembled from three equally sized achiral nanodisks, is super-imposable with its mirror-image by a simple rotation in two-dimensional space (see Fig. \ref{fig:Fig1}a). In contrast, the trimer is not super-imposable with its mirror image anymore if the heterogeneous material composition is taken into account (see Figs. \ref{fig:Fig1}b and \ref{fig:Fig1}c). As we will show later on, it is therefore intrinsically chiral. 

\section*{Results}
\subsection{Numerical Results for trimers built from nanodisks}
For our numerical study we have chosen nanodisks made from gold (Au) and silicon (Si). The spectra of individual Si nanodisks exhibit multiple resonances depending on the particle size, with the magnetic dipole representing the fundamental particle resonance. It becomes apparent that this behavior is caused by the dielectric properties and the high refractive index of Si nanoparticles in the visible and near-IR spectral range that has been studied extensively in the recent years\cite{Evlyukhin2010,Garcia2011,Kuznetsov2012,vandeGroep2013,Wozniak2015}. The nanodisks have a diameter of 180 nm and a height of 30 nm embedded into a homogeneous environment with a refractive index of $n_{air}$ = 1 (see Figs. \ref{fig:Fig1}b and \ref{fig:Fig1}c). For excitation, we illuminate the trimer with a weakly focused Gaussian light beam (numerical aperture of NA = 0.4) propagating along the negative $z$-direction. This choice is also motivated by the fact that for this low NA, the structures are excited by a dominantly transverse electric field (plane-wave-like). The generation of longitudinal electric field components (which carry a phase vortex) upon tight focusing and their additional interaction with the trimer under study is therefore minimized\cite{Zhao2007}. It should be emphasized here again that we aim for studying individual trimeric structures only because we are interested in their intrinsic chiral optical response rather than collective effects caused by an arrangement of such structures in two-dimensional arrays or metasurfaces.\\ 

\begin{figure}
\includegraphics[width=0.8\textwidth]{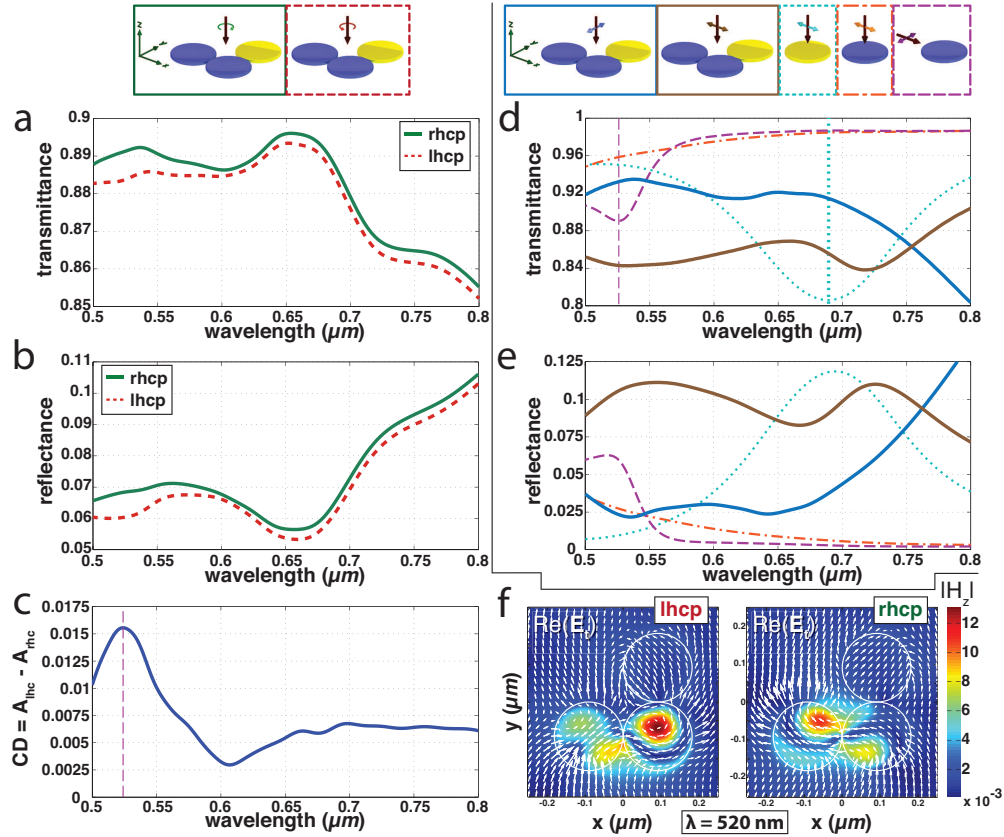}
\caption{\label{fig:Fig2}\textbf{Numerical results for different nanodisks system}. Numerically calculated (a) transmittance and (b) reflectance spectra for an individual nanodisk trimer for left-handed circularly polarized (lhcp, dashed red line) and right-handed circularly polarized (rhcp; solid green line) light excitation. (c) Resulting circular dichroism curve. (d) Transmittance and (e) reflectance spectra for a trimer (solid blue and brown curves) and individual nanodisks (dotted and dashed curves) for linearly polarized excitation. The corresponding excitation schemes are indicated above (d). The vertical dotted cyan-colored line indicates the spectral position of the electric dipole resonance of an individual Au nanodisk, whereas the vertical dashed magenta-colored line marks the position of the magnetic dipole resonance of an individual Si nanodisk for the indicated excitation schemes. (f) Distributions of the instantaneous in-plane electric field ($\mathbf{E}_{t} = \left(E_{x},E_{y}\right)$; white arrows) and the modulus of the longitudinal magnetic field component ($\left|H_{z}\right|$; color-coded) for a wavelength of 520 nm plotted in a plane parallel to the $xy$-plane cutting through the centers of the disks.}
\end{figure}

First, we present the spectra of individual trimers embedded in air for right- and left-handed circularly polarized (lhcp, rhcp) excitation under normal incidence to underline the statement made above that individual trimers with a material-composition-induced break of symmetry exhibit chiral properties (see transmittance in Fig. \ref{fig:Fig2}a and reflectance in Fig. \ref{fig:Fig2}b). The CD spectrum of the structure can be calculated as $CD  =  A_{lhcp} - A_{rhcp}$ where $ A = 1 - T - R$ is the absorbance, and $T$ and $R$ are the transmittance and reflectance, respectively. The CD spectrum exhibits a maximum at a wavelength of 520 nm and is positive throughout the investigated spectrum, indicating a fixed optical handedness of the structure in this range (see Fig. \ref{fig:Fig2}c). By changing the order of the nanodisks in the trimer (exchanging the first Si disk (1) with the Au disk), the handedness of the trimer is inverted and the CD spectrum changes its sign (see also Supplementary Fig. 1).\\
To understand the origin of this CD spectrum, we study the response of individual nanodisks. The resonance spectra of Au nanodisks for linearly polarized excitation (Figs. \ref{fig:Fig2}d and \ref{fig:Fig2}e; dotted cyan-colored curve) are dominated by the electric dipole mode (vertical dotted line). On the other hand, the Si nanodisk exhibits no resonant features in the spectral range studied here (see also Figs. \ref{fig:Fig2}d and \ref{fig:Fig2}e; dotted-dashed red curve) when excited with linearly polarized light from the top.  It can be shown that this is a simple consequence of the chosen height of the disk\cite{vandeGroep2013}. In contrast, the situation is considerably different for an excitation of the disk with light being linearly polarized and propagating in the $xy$-plane. This scenario mimics the case of light scattering in the plane of the nanodisks and exciting neighboring particles in the trimer scheme. The corresponding spectra (see again Figs. \ref{fig:Fig2}d and \ref{fig:Fig2}e; dashed magenta-colored curve) show a clear resonance at a wavelength of 520 nm (vertical dashed line). A detailed study of the electric and magnetic fields in the disk forming the resonant mode reveals that it can be associated with the fundamental magnetic dipole resonance formed by a curl of the electric field inside the $xy$-plane inside disk. This resonant mode cannot be excited directly by a linearly or circularly polarized light beam impinging along the $z$-axis, but it may be induced by the mutual interaction between neighboring particles in the trimer.\\

For the sake of completeness, we also include the resonance spectra of the heterogeneously composed trimer for $x$- ($E_{x,inc}$) and $y$-polarized ($E_{y,inc}$) excitation in Figs. \ref{fig:Fig2}d and \ref{fig:Fig2}e. The spectra for both excitations schemes differ drastically from each other, also indicating the material-induced asymmetry of the studied system.\\
In the trimeric arrangement, the transverse dipole moments of the individual disks can be excited (non-resonantly) both directly by the impinging circularly polarized light field and by light scattered off neighboring particles. This leads to hybridization of the modes\cite{Albella2013,Bakker2015}. It is this hybridization of the disk modes for the excitation with $x$- and $y$-polarized light that leads to different resonance spectra (see Figs. \ref{fig:Fig2}d and \ref{fig:Fig2}e), due to the chosen material-induced break of symmetry. In addition, the relative phase between the transverse electric field components $E_{x,inc}$ and $E_{y,inc}$ for circularly polarized excitation ($\pm \pi/2$) defines the phase between the induced transversely oriented dipole moments. The key to understanding the emerging optical chirality of the nanostructure at hand is the interaction between the modes excited directly by the incoming light field with the same modes excited via particle scattering or hybridization. Furthermore, also the interference between additional particle modes contributes, induced, for instance, in the Si disks solely by scattering of neighboring particles for normal excitation. This phenomenon can be seen in Fig. \ref{fig:Fig2}f, where the field distributions in a cut-plane through the trimeric disk assembly are plotted for both handedness of the excitation field and at a fixed wavelength of 520 nm, for which the CD spectra peaked. The plotted fields exhibit different spatial distributions indicating that for both cases different particle modes contribute to the scattering process. Both, the plotted distributions of the instantaneous electric field (white arrows; transverse components shown only) and the modulus of the longitudinal magnetic field component ($\left|H_{z}\right|$) indicate that for left-handed circularly polarized excitation (Fig. \ref{fig:Fig2}f, left), a longitudinal magnetic dipole component is observable dominating the mode in the central Si particle (2). Consequently, the coupling of the particles with each other induces an additional dipole contribution. In the other Si disk (1), the mode is more complex. For right-handed circularly polarized light (Fig. \ref{fig:Fig2}f, right), the situation is different. The mode structure in the central Si disk is also more complex and does not exhibit a clear fingerprint of a longitudinal magnetic dipole moment anymore.\\
\begin{figure}
\includegraphics[width=0.9\textwidth]{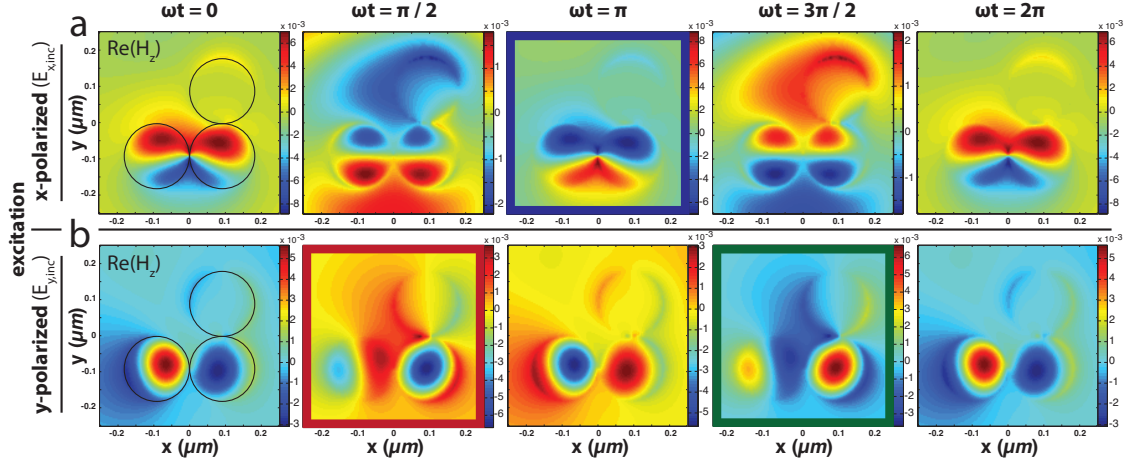}
\caption{\label{fig:Fig2a}\textbf{Near-field maps for different excitation schemes.} Time-evolution of the longitudinal magnetic field component Re($H_{z}$) plotted in a plane parallel to the $xy$-plane cutting through the centers of all particles for two different excitation schemes and a fixed wavelength of 520 nm. (a) Excitation with an $x$-polarized Gaussian light beam. (b) Excitation with a $y$-polarized Gaussian light beam. Each field map is normalized to its individual minima and maxima resulting in different color-scales for different maps. The geometrical outlines of the three nanosdisks (top: Au; bottom: Si) are indicated by black circles. For right-handed (left-handed) circularly polarized excitation the frames indicated in blue in (a) and green (red) in b overlap temporally.}
\end{figure} 
Not less important for this discussion is the fact that even though also the same particle modes (transverse dipoles) might be involved in the optical response of the trimer for both handednesses of excitation, they will differ in their multipole amplitudes due to the presence of the Au disk and, hence, contribute with different strengths. Ultimately, those modes excited by scattering can be different for the $x$- and $y$-field components of the impinging circularly polarized light beam (see Fig. \ref{fig:Fig2a}), due to the symmetry and geometry of the trimer. This causes an additional phase delay between the directly excited and scattering-induced particle modes. They can, therefore, interfere with each other coherently, resulting in a suppression or enhancement of such particle modes, which ultimately depends on the relative phase between of the components of the excitation field or, equivalently, on the handedness of the excitation.\\
The aforementioned arguments become more apparent when looking at the induced field patterns for $x$- and $y$-polarized excitation as shown exemplarily in Fig. \ref{fig:Fig2a} again for a wavelength of 520 nm. This field-based analysis also enables a more quantitative discussion of the material-composition-induced chirality proposed here. In Fig. \ref{fig:Fig2a}a, the time evolution of the instantaneous longitudinal magnetic field component ($H_{z}$) is plotted in steps of $\pi/2$ for an excitation with a x-polarized Gaussian beam. For comparison, the corresponding numerical results for the same trimer excited with $y$-polarized light are shown in Fig. \ref{fig:Fig2a}b. For $x$-polarized excitation, the field distributions inside the Si nanodisks resemble a two-lobe pattern, which is a characteristic fingerprint of the field created by an electric dipole mode oscillating along the $x$-axis. The actual electric dipole resonances of the individual Si nanodisks are found outside the investigated spectral range. Hence, the excited mode is driven off-resonance. The distributions are altered by the near-field interaction of both particles, indicating mode hybridization. As time evolves, the magnetic field oscillates, preserving the initial distributions of the field structure throughout a full period, which are similar and in-phase for both Si disks. For symmetry reasons, the scattering and near-fields of the Si particle at the lower left position (1) of the trimer cannot efficiently excite a resonant longitudinal magnetic dipole moment (along the $z$-axis) in the second Si particle or vice versa, because its field is dominantly polarized along the Si-Si-dimer axis. In this context, the contribution of the scattering off the Au particle to the field structure is minor, because it is driven far off-resonance.\\
In contrast to the excitation with $x$-polarization ($\mathbf{E}_{inc} = \left(E_{x,inc}, 0, 0\right)$), the field patterns in both Si disks differ strongly when excited with $y$-polarized light ($\mathbf{E}_{inc} = \left(0, E_{y,inc}, 0\right)$) (see Fig. \ref{fig:Fig2a}b). For this case, the field distribution inside the central Si particle (2) exhibits only one strong and dominant lobe, which oscillates in time parallel to the $z$-axis. The corresponding electric field shows a curl-like distribution. This field pattern corresponds to that of a longitudinally oscillating magnetic dipole moment, which, for an individual disk, is resonantly excited at the chosen wavelength (see Figs. \ref{fig:Fig2}d and \ref{fig:Fig2}e). It is induced by the vertically ($y$) oscillating transverse electric dipole mode in the neighboring Si disk (1), excited directly by the incoming light field (see, for instance, second image in Fig \ref{fig:Fig2a}b). The same holds true for the vertically oriented electric dipole moment excited in the central Si particle (2), inducing a magnetic dipole moment in the left-hand Si disk (1). Nonetheless, the time evolution of the fields within both disks is not the same, which is a consequence of the presence of the Au particle and, thus, a result of the material-induced break of symmetry. It should be mentioned here, that beside the coupling of the longitudinal magnetic to the transverse electric dipole moments of the Si disks, also the electric dipole moments of both Si nanodisks themselves are coupled to each other\cite{Albella2013}.\\
After this analysis it becomes clear, why the field distributions for right- and left-handed circularly polarized excitation as shown in Fig. \ref{fig:Fig2}f differ from each other drastically and why they are not symmetric, as they would be for a simple Si dimer. For right-handed and left-handed circularly polarized light, the incoming beam can be written as $\mathbf{E}_{inc} = \left(E_{x,inc}, i  E_{y,inc}, 0\right)$ and $\mathbf{E}_{inc} = \left(E_{x,inc}, -i  E_{y,inc}, 0\right)$, respectively. As a consequence of the excitation with circularly polarized light, the two temporal field evolutions shown in Figs. \ref{fig:Fig2a}a and \ref{fig:Fig2a}b have to be shifted relative to each other due to an additional relative phase-delay between the corresponding field distributions of $\mp \pi/2$. As a result, the frames indicated in blue in Fig. \ref{fig:Fig2a}a and green in Fig. \ref{fig:Fig2a}b overlap temporally for right-handed circularly polarized excitation, whereas the blue frame overlaps with the red frame for left-hand circular polarization. The corresponding field distributions in Figs. \ref{fig:Fig2a}a and \ref{fig:Fig2a}b therefore interfere with each other forming the patterns shown in \ref{fig:Fig2}f. The observed asymmetries between the near-fields of the individual heterogeneous trimer are consequently a result of constructive and destructive interference of the symmetric and asymmetric field patterns in Figs. \ref{fig:Fig2a}a and \ref{fig:Fig2a}b. Ultimately, this interference results in different resonance spectra for excitation with different handedness of the incoming light.

\subsection{Experimental results for trimers built from nanospheres}
\begin{figure}
\includegraphics[width=0.5\textwidth]{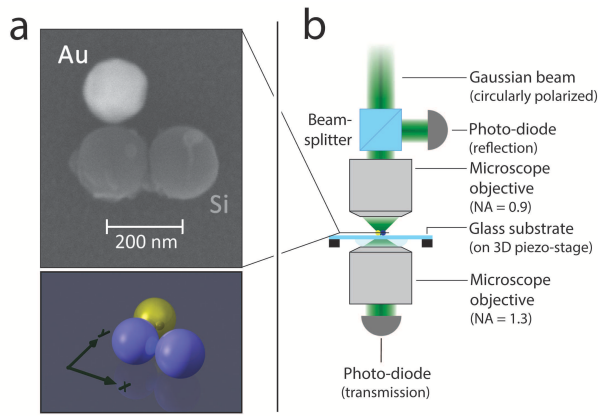}
\caption{\label{fig:Fig3}\textbf{Investigated nanostructure and experimental setup.} (a) Top: SEM micrograph of the investigated trimer assembled from one Au (white; diameter $\approx$ 175 nm) and two Si (dark gray; diameters $\approx$ 180 nm) nanospheres on a borosilicate glass (BK7) substrate. Bottom: Sketch of the trimer and chosen coordinate frame. It should be noted here that the investigated trimer exhibits a mirrored geometry with respect to the nanodisk trimer discussed before. (b) Simplified sketch of the experimental setup utilized for the measurement of individual nanosphere trimers. A circularly polarized fundamental Gaussian light beam enters a microscope objective (MO) with a high numerical (NA) of 0.9. The entrance aperture of the MO is not filled, hence reducing the effective NA to approximately 0.4. The focused light beam has a diameter of approximately 0.8 $\mu$m (FWHM) at a wavelength of 625 nm. The sample composed of a substrate with individual nanotrimers assembled on top is mounted on a 3D-piezo-stage, enabling precise positioning of the structure relative to the beam. A second immersion-type MO with an NA of 1.3 is located below the sample to collect the light in the forward direction, which is detected with a photo-diode. Light reflected or scattered backwards is collected by the upper MO and guided to a second photo-diode by a set of non-polarizing beamsplitters.}
\end{figure}

\begin{figure}
\includegraphics[width=0.9\textwidth]{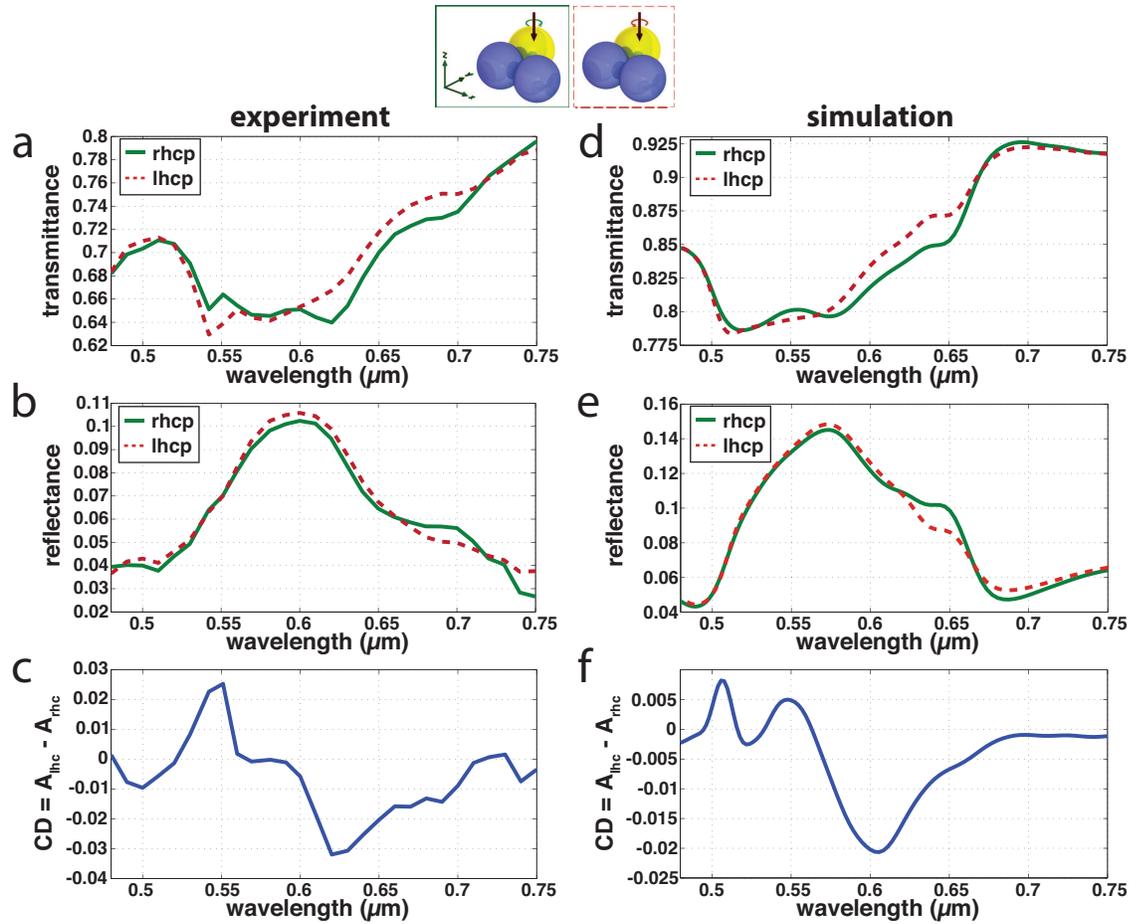}
\caption{\label{fig:Fig4}\textbf{Comparison of experimental and numerical results for a heterogeneous nanosphere trimer.} (a) - (c) Experimental and (d) - (e) numerical results for an individual trimer assembled from Au and Si nanospheres on a glass substrate (see Fig. \ref{fig:Fig3}). Transmittance and reflectance as well as the CD spectra are shown. In the experiment, a thin layer of AZO was present on top of the glass substrate to prevent charging effects during fabrication.}
\end{figure}

For the experimental demonstration of the proposed scheme, we choose a nanotrimer assembled from individual nanospheres (see Fig. \ref{fig:Fig3}a) rather than nanodisks as studied numerically above. We use a pick-and-place-based approach for the assembly of spherical nanoparticles, realized by operating an atomic force microscope (AFM) inside the vacuum-chamber of a scanning electron microscope (SEM)\cite{Bartenwerfer2011,Mick2014} (see also Methods section). Because we are interested in the optical measurement of a single trimer only, the pick-and-place-based fabrication is straight-forward and time-efficient. A fabrication of trimers consisting of nanodisks rather than nanospheres would require multiple complex and time-consuming coating and lithography steps, which is very inconvenient for the investigation of the proposed scheme and the proof of its principle. The scheme involving nanodisks as discussed above enabled a more fundamental discussion and intuitive understanding of the effects causing a chiral behavior of heterogeneous geometrically symmetric trimers. Nonetheless, the underlying physics are similar for both systems (disks and spheres).\\
To be able to study an individual nanotrimer experimentally, we utilize a custom-built optical setup\cite{Banzer2010a}. A sketch of the system is shown in Fig. \ref{fig:Fig3}b (see Methods section for details). For the measurements, the wavelength is selected with the spectral filter, the nanotrimer is placed in the focal plane and it is scanned line-by-line through the focal spot several times. From such measurements, the transmitted and reflected power values for the particle sitting effectively on the optical axis in the focal plane are retrieved and averaged. The corresponding data is normalized to the power transmitted through and reflected from the air-substrate interface, respectively. Following this procedure, the measurements are repeated for different wavelengths and for both handednesses of the circularly polarized input beam, resulting in the corresponding normalized resonance spectra in the spectral range from 480 to 750 nm as shown in Figs. \ref{fig:Fig4}a and \ref{fig:Fig4}b. The figure also includes the simulation results for the investigated scheme (see Figs. \ref{fig:Fig4}d and \ref{fig:Fig4}e), taking into account the experimental conditions, such as the experimental arrangement of the individual particles, their approximate sizes (outer diameter of 180 nm), the BK7 substrate and the beam configuration. Furthermore, we also incorporate a thin SiO$_2$ surface layer around the Si nanospheres naturally occurring due to oxidation. The layer thickness is estimated to be approximately 8 nm\cite{Wozniak2015}. The actual diameter of the Si core is therefore reduced correspondingly. In the experiment, a thin layer of AZO between the substrate and the trimer was used to suppress charging effects during the pick-and-place fabrication. This layer is not taken into account in the simulations. The experimental data is normalized to the transmission through and reflection at the AZO-on-glass substrate, whereas the numerical data was retrieved and normalized for a glass substrate only.\\
From both sets of data, we calculate the spectral distributions of the resulting CD spectrum, displayed in Figs. \ref{fig:Fig4}c and \ref{fig:Fig4}f for the experimental and numerical results, respectively. The experimental reflectance and transmittance curves in Figs. \ref{fig:Fig4}a and \ref{fig:Fig4}b bear striking similarities with the spectra retrieved from numerical calculations (see Figs. \ref{fig:Fig4}d and \ref{fig:Fig4}e). The slight red-shift of the experimental data with respect to the numerical results and also the different absolute values for transmittance and reflectance data sets are caused by the presence of the aforementioned AZO-layer on the experimental sample. Nonetheless, both datasets underline the chiral behavior of the geometrically mirror-symmetric particle ensemble (trimer). Apparently, these spectra are more rich in comparison to those of trimers composed of nanodisks discussed above. This is caused by the fact that for nanospheres of the chosen size made from high refractive index dielectrics such as Si in the visible or near-infrared spectral range, magnetic and electric dipole resonances appear at wavelengths in the investigated spectral range. Therefore, the incoming light field can now also resonantly excite a magnetic dipole -- a transversely oscillating magnetic dipole moment perpendicular to the $z$-axis -- in the investigated spectral range, which contributes to the resonance behavior of the trimer. Furthermore, the utilized substrate as a support for the particles or, in other words, the air-dielectric interface also influences the retrieved spectra. As discussed before, the substrate additionally breaks the forward-backward symmetry of the system, further influencing its chiral response. Nonetheless, the trimer assembled from nanospheres as investigated here of course also shows a chiral response if embedded in a homogeneous environment. It is also worth noting that the corresponding CD spectra for the investigated trimer made from nanospheres show a change of sign in certain spectral regions. Hence, the actual optical handedness of the trimer also changes and, therefore, depends on the wavelength. For comparison, we also assembled and measured a nanotrimer with opposite handedness resulting in an inversion of the CD spectra (see Supplementary Fig. 1).\\
The observed CD spectra can be further modified easily by assembling trimers from smaller or larger nanopheres, by changing the size of the central particle relative to the other particles, still keeping the trimer geometrically symmetric, or by using nanospheres made from  materials different than Au and Si. These approaches allow for tuning the particle spectra relative to each other. For instance for larger particles, also higher order resonances contribute to the scattering and the inter-particle coupling, enhancing the performance of the chiral particle ensemble. It is also conceivable, that the trimer may be composed of equally sized particles of three different materials or from more than just three particles. These design criteria allow for a broad range of structural arrangements, further tuning and boosting the material-composition-induced chiral effect discussed above.\\

In conclusion, we have introduced and studied -- as a proof-of-principle -- a novel approach for the realization of optically chiral nanostructures. The proposed scheme is based on a geometrically achiral nanoparticle (sphere, disks or similar) assembly, which nonetheless exhibits a chiral optical response due to its heterogeneous composition. The chiral behavior of the trimer is caused by the mutual interaction of neighboring particles when excited under normal incidence. To demonstrate this phenomenon also experimentally, we have fabricated heterogeneous trimers from individual Au and Si nanospheres on a substrate, using a highly precise pick-and-place technique realized by operating an AFM inside the vacuum chamber of an SEM. Such heterogeneously-composed particle ensembles allow for a high tunability with respect to their spectra and chiral optical response. With the possibility of tailoring the multipolar response of the individual constituents, the response of the chiral assemblies can be tuned in a straight-forward fashion over a broad spectral range. Furthermore, we believe that based on this novel material composition-enabled approach to chirality of nanostructures introduced and discussed in this Letter, also large-scale chiral materials and metasurfaces can be designed to allow for the realization of ultra-thin chiral filters and polarizers. Potential approaches for fabricating large arrays of individual heterogeneous nanoparticle assemblies could be based on the methods recently described in literature\cite{Yang2014,Kaell2014}. It is worth noting that based on the arguments discussed above, some of the trimeric particle arrangements presented in the corresponding reference \citen{Yang2014} should also show a chiral optical behavior.

\section*{Methods}
\subsection{Sample fabrication}
For the fabrication of individual heterogeneously composed trimers we apply a pick-and-place-based manipulation approach. This method is based on operating an atomic force microscope (AFM) inside the vacuum-chamber of a scanning electron microscope (SEM)\cite{Bartenwerfer2011,Mick2014}. In this custom-built system, the AFM tip is modified to permit precise handling of nanoparticles. This approach allows for choosing particles with the desired size, shape and material from a particle reservoir and enables precise positioning of the individual particles relative to each other on a substrate. To prevent charging of the substrate resulting from the imaging process of the sample with the electron beam while manipulating the nanoparticles, a thin AZO-layer is deposited on the borosilicate glass (BK7) substrate first.

\subsection{Experimental setup and measurement scheme}
The custom-built optical setup (see Fig. \ref{fig:Fig3}b) consists of a spectral filter (acousto-optical spectral filter; spectral bandwidth: 1 nm) placed behind the output of a broadband fiber-based light source. It selects the central wavelength of the transmitted Gaussian light beam. The light beam is transmitted through a quarter-wave plate. The resulting circularly polarized beam is guided top-down into a microscope objective with a high numerical aperture (NA) and focused onto the sample. The size of the input beam is chosen such that the input beam does not fill the entrance aperture of the microscope objective. This way, an effectively smaller focusing NA (0.4) is achieved and the influence of longitudinal electric field components created upon strong focusing is reduced\cite{Zhao2007}. The sample is composed of a borosilicate glass (BK7) substrate with a thickness of approximately 170 $\mu$m and a thin AZO-layer on top. On the surface of the substrate, individual trimeric structures are assembled using the above-mentioned pick-and-place handling technique. Fig. \ref{fig:Fig3}a shows an SEM micrograph of the assembly investigated here. The sample is mounted on a piezo-stage, which allows for precisely positioning it relative to the excitation beam. Light transmitted through the substrate or scattered in the forward direction by the trimer is collected by a second microscope objective. This oil-immersion objective with an NA of 1.3 collimates the collected light. The resulting beam is guided to a photodetector. Light reflected from the air-substrate interface or back-scattered light is collected within a large solid angle by the upper objective, originally used for focusing the incoming light beam. A non-polarizing beamsplitter inserted in front of this focusing objective allows for separating the incoming and a part of the reflected light beam and guides the latter to a detector. For guaranteeing a high quality of the input and the reflected light beams with respect to their polarization, we utilize a set of beam-splitters, all non-polarizing in type, for preserving the polarization state\cite{Tidwell1992,ZhanBook2014}.\\

\bibliography{manuscript_chiral_nanotrimer_draft_final}

\begin{thebibliography}{10}
\expandafter\ifx\csname url\endcsname\relax
  \def\url#1{\texttt{#1}}\fi
\expandafter\ifx\csname urlprefix\endcsname\relax\def\urlprefix{URL }\fi
\providecommand{\bibinfo}[2]{#2}
\providecommand{\eprint}[2][]{\url{#2}}

\bibitem{Gansel2009}
\bibinfo{author}{Gansel, J.~K.} \emph{et~al.}
\newblock \bibinfo{title}{Gold helix photonic metamaterial as broadband
  circular polarizer}.
\newblock \emph{\bibinfo{journal}{Science}} \textbf{\bibinfo{volume}{325}},
  \bibinfo{pages}{1513--1515} (\bibinfo{year}{2009}).

\bibitem{Thiel2010}
\bibinfo{author}{Thiel, M.}, \bibinfo{author}{Fischer, H.},
  \bibinfo{author}{von Freymann, G.} \& \bibinfo{author}{Wegener, M.}
\newblock \bibinfo{title}{Three-dimensional chiral photonic superlattices}.
\newblock \emph{\bibinfo{journal}{Opt. Lett.}} \textbf{\bibinfo{volume}{35}},
  \bibinfo{pages}{166--168} (\bibinfo{year}{2010}).

\bibitem{Hoeflich2011}
\bibinfo{author}{H{\"o}flich, K.}, \bibinfo{author}{Yang, R.~B.},
  \bibinfo{author}{Berger, A.}, \bibinfo{author}{Leuchs, G.} \&
  \bibinfo{author}{Christiansen, S.}
\newblock \bibinfo{title}{The direct writing of plasmonic gold nanostructures
  by electron-beam-induced deposition}.
\newblock \emph{\bibinfo{journal}{Adv. Mater.}} \textbf{\bibinfo{volume}{23}},
  \bibinfo{pages}{2657--2661} (\bibinfo{year}{2011}).

\bibitem{Helgert2011}
\bibinfo{author}{Helgert, C.} \emph{et~al.}
\newblock \bibinfo{title}{Chiral metamaterial composed of three-dimensional
  plasmonic nanostructures}.
\newblock \emph{\bibinfo{journal}{Nano Lett.}} \textbf{\bibinfo{volume}{11}},
  \bibinfo{pages}{4400--4404} (\bibinfo{year}{2011}).

\bibitem{Frank2013}
\bibinfo{author}{Frank, B.} \emph{et~al.}
\newblock \bibinfo{title}{Large-area 3d chiral plasmonic structures}.
\newblock \emph{\bibinfo{journal}{ACS Nano}} \textbf{\bibinfo{volume}{7}},
  \bibinfo{pages}{6321--6329} (\bibinfo{year}{2013}).

\bibitem{Belardini2014}
\bibinfo{author}{Belardini, A.} \emph{et~al.}
\newblock \bibinfo{title}{Second harmonic generation circular dichroism from
  self-ordered hybrid plasmonic–photonic nanosurfaces}.
\newblock \emph{\bibinfo{journal}{Adv. Opt. Mater.}}
  \textbf{\bibinfo{volume}{2}}, \bibinfo{pages}{208--213}
  (\bibinfo{year}{2014}).

\bibitem{Esposito2015}
\bibinfo{author}{Esposito, M.} \emph{et~al.}
\newblock \bibinfo{title}{Nanoscale 3d chiral plasmonic helices with circular
  dichroism at visible frequencies}.
\newblock \emph{\bibinfo{journal}{ACS Photonics}} \textbf{\bibinfo{volume}{2}},
  \bibinfo{pages}{105--114} (\bibinfo{year}{2015}).

\bibitem{Schaeferling2012}
\bibinfo{author}{Sch{\"a}ferling, M.}, \bibinfo{author}{Dregely, D.},
  \bibinfo{author}{Hentschel, M.} \& \bibinfo{author}{Giessen, H.}
\newblock \bibinfo{title}{Tailoring enhanced optical chirality: Design
  principles for chiral plasmonic nanostructures}.
\newblock \emph{\bibinfo{journal}{Phys. Rev. X}} \textbf{\bibinfo{volume}{2}},
  \bibinfo{pages}{031010} (\bibinfo{year}{2012}).

\bibitem{Valev2013}
\bibinfo{author}{Valev, V.~K.}, \bibinfo{author}{Baumberg, J.~J.},
  \bibinfo{author}{Sibilia, C.} \& \bibinfo{author}{Verbiest, T.}
\newblock \bibinfo{title}{Chirality and chiroptical effects in plasmonic
  nanostructures: Fundamentals, recent progress, and outlook}.
\newblock \emph{\bibinfo{journal}{Adv. Mater.}} \textbf{\bibinfo{volume}{25}},
  \bibinfo{pages}{2517--2534} (\bibinfo{year}{2013}).

\bibitem{Tang2010}
\bibinfo{author}{Tang, Y.} \& \bibinfo{author}{Cohen, A.~E.}
\newblock \bibinfo{title}{Optical chirality and its interaction with matter}.
\newblock \emph{\bibinfo{journal}{Phys. Rev. Lett.}}
  \textbf{\bibinfo{volume}{104}}, \bibinfo{pages}{163901}
  (\bibinfo{year}{2010}).

\bibitem{Bliokh2011}
\bibinfo{author}{Bliokh, K.~Y.} \& \bibinfo{author}{Nori, F.}
\newblock \bibinfo{title}{Characterizing optical chirality}.
\newblock \emph{\bibinfo{journal}{Phys. Rev. A}} \textbf{\bibinfo{volume}{83}},
  \bibinfo{pages}{021803} (\bibinfo{year}{2011}).

\bibitem{Coles2012}
\bibinfo{author}{Coles, M.~M.} \& \bibinfo{author}{Andrews, D.~L.}
\newblock \bibinfo{title}{Chirality and angular momentum in optical radiation}.
\newblock \emph{\bibinfo{journal}{Phys. Rev. A}} \textbf{\bibinfo{volume}{85}},
  \bibinfo{pages}{063810} (\bibinfo{year}{2012}).

\bibitem{Fernandez2013}
\bibinfo{author}{Fernandez-Corbaton, I.}, \bibinfo{author}{Vidal, X.},
  \bibinfo{author}{Tischler, N.} \& \bibinfo{author}{Molina-Terriza, G.}
\newblock \bibinfo{title}{Necessary symmetry conditions for the rotation of
  light}.
\newblock \emph{\bibinfo{journal}{J. Chem. Phys.}}
  \textbf{\bibinfo{volume}{138}}, \bibinfo{pages}{--} (\bibinfo{year}{2013}).

\bibitem{Liu2007}
\bibinfo{author}{Liu, H.} \emph{et~al.}
\newblock \bibinfo{title}{Magnetic plasmon hybridization and optical activity
  at optical frequencies in metallic nanostructures}.
\newblock \emph{\bibinfo{journal}{Phys. Rev. B}} \textbf{\bibinfo{volume}{76}},
  \bibinfo{pages}{073101} (\bibinfo{year}{2007}).

\bibitem{Liu2009}
\bibinfo{author}{Liu, N.}, \bibinfo{author}{Liu, H.}, \bibinfo{author}{Zhu, S.}
  \& \bibinfo{author}{Giessen, H.}
\newblock \bibinfo{title}{Stereometamaterials}.
\newblock \emph{\bibinfo{journal}{Nat. Photon.}} \textbf{\bibinfo{volume}{3}},
  \bibinfo{pages}{157--162} (\bibinfo{year}{2009}).

\bibitem{Decker2009}
\bibinfo{author}{Decker, M.} \emph{et~al.}
\newblock \bibinfo{title}{Strong optical activity from twisted-cross photonic
  metamaterials}.
\newblock \emph{\bibinfo{journal}{Opt. Lett.}} \textbf{\bibinfo{volume}{34}},
  \bibinfo{pages}{2501--2503} (\bibinfo{year}{2009}).

\bibitem{Fan2010}
\bibinfo{author}{Fan, Z.} \& \bibinfo{author}{Govorov, A.~O.}
\newblock \bibinfo{title}{Plasmonic circular dichroism of chiral metal
  nanoparticle assemblies}.
\newblock \emph{\bibinfo{journal}{Nano Lett.}} \textbf{\bibinfo{volume}{10}},
  \bibinfo{pages}{2580--2587} (\bibinfo{year}{2010}).

\bibitem{Kuzyk2012}
\bibinfo{author}{Kuzyk, A.} \emph{et~al.}
\newblock \bibinfo{title}{Dna-based self-assembly of chiral plasmonic
  nanostructures with tailored optical response}.
\newblock \emph{\bibinfo{journal}{Nature}} \textbf{\bibinfo{volume}{483}},
  \bibinfo{pages}{311--314} (\bibinfo{year}{2012}).

\bibitem{Hentschel2012}
\bibinfo{author}{Hentschel, M.}, \bibinfo{author}{Sch{\"a}ferling, M.},
  \bibinfo{author}{Metzger, B.} \& \bibinfo{author}{Giessen, H.}
\newblock \bibinfo{title}{Plasmonic diastereomers: Adding up chiral centers}.
\newblock \emph{\bibinfo{journal}{Nano Lett.}} \textbf{\bibinfo{volume}{13}},
  \bibinfo{pages}{600--606} (\bibinfo{year}{2013}).

\bibitem{Shen2013}
\bibinfo{author}{Shen, X.} \emph{et~al.}
\newblock \bibinfo{title}{Three-dimensional plasmonic chiral tetramers
  assembled by dna origami}.
\newblock \emph{\bibinfo{journal}{Nano Lett.}} \textbf{\bibinfo{volume}{13}},
  \bibinfo{pages}{2128--2133} (\bibinfo{year}{2013}).

\bibitem{Kaell2014}
\bibinfo{author}{Ogier, R.}, \bibinfo{author}{Fang, Y.},
  \bibinfo{author}{Svedendahl, M.}, \bibinfo{author}{Johansson, P.} \&
  \bibinfo{author}{K{\"a}ll, M.}
\newblock \bibinfo{title}{Macroscopic layers of chiral plasmonic nanoparticle
  oligomers from colloidal lithography}.
\newblock \emph{\bibinfo{journal}{ACS Photonics}} \textbf{\bibinfo{volume}{1}},
  \bibinfo{pages}{1074--1081} (\bibinfo{year}{2014}).

\bibitem{Papakostas2003}
\bibinfo{author}{Papakostas, A.} \emph{et~al.}
\newblock \bibinfo{title}{Optical manifestations of planar chirality}.
\newblock \emph{\bibinfo{journal}{Phys. Rev. Lett.}}
  \textbf{\bibinfo{volume}{90}}, \bibinfo{pages}{107404}
  (\bibinfo{year}{2003}).

\bibitem{Rogacheva2006}
\bibinfo{author}{Rogacheva, A.~V.}, \bibinfo{author}{Fedotov, V.~A.},
  \bibinfo{author}{Schwanecke, A.~S.} \& \bibinfo{author}{Zheludev, N.~I.}
\newblock \bibinfo{title}{Giant gyrotropy due to electromagnetic-field coupling
  in a bilayered chiral structure}.
\newblock \emph{\bibinfo{journal}{Phys. Rev. Lett.}}
  \textbf{\bibinfo{volume}{97}}, \bibinfo{pages}{177401}
  (\bibinfo{year}{2006}).

\bibitem{Reichelt2006}
\bibinfo{author}{Reichelt, M.} \emph{et~al.}
\newblock \bibinfo{title}{Broken enantiomeric symmetry for electromagnetic
  waves interacting with planar chiral nanostructures}.
\newblock \emph{\bibinfo{journal}{Appl. Phys. B}}
  \textbf{\bibinfo{volume}{84}}, \bibinfo{pages}{97--101}
  (\bibinfo{year}{2006}).

\bibitem{Decker2007}
\bibinfo{author}{Decker, M.}, \bibinfo{author}{Klein, M.~W.},
  \bibinfo{author}{Wegener, M.} \& \bibinfo{author}{Linden, S.}
\newblock \bibinfo{title}{Circular dichroism of planar chiral magnetic
  metamaterials}.
\newblock \emph{\bibinfo{journal}{Opt. Lett.}} \textbf{\bibinfo{volume}{32}},
  \bibinfo{pages}{856--858} (\bibinfo{year}{2007}).

\bibitem{Plum2009a}
\bibinfo{author}{Plum, E.}, \bibinfo{author}{Fedotov, V.~A.} \&
  \bibinfo{author}{Zheludev, N.~I.}
\newblock \bibinfo{title}{Extrinsic electromagnetic chirality in
  metamaterials}.
\newblock \emph{\bibinfo{journal}{J. Opt. A: Pure Appl. Opt.}}
  \textbf{\bibinfo{volume}{11}}, \bibinfo{pages}{074009}
  (\bibinfo{year}{2009}).

\bibitem{Hendry2010}
\bibinfo{author}{Hendry, E.} \emph{et~al.}
\newblock \bibinfo{title}{Ultrasensitive detection and characterization of
  biomolecules using superchiral fields}.
\newblock \emph{\bibinfo{journal}{Nature Nanotech.}}
  \textbf{\bibinfo{volume}{5}}, \bibinfo{pages}{783--787}
  (\bibinfo{year}{2010}).

\bibitem{Zhao2011}
\bibinfo{author}{Zhao, R.}, \bibinfo{author}{Zhang, L.}, \bibinfo{author}{Zhou,
  J.}, \bibinfo{author}{Koschny, T.} \& \bibinfo{author}{Soukoulis, C.~M.}
\newblock \bibinfo{title}{Conjugated gammadion chiral metamaterial with
  uniaxial optical activity and negative refractive index}.
\newblock \emph{\bibinfo{journal}{Phys. Rev. B}} \textbf{\bibinfo{volume}{83}},
  \bibinfo{pages}{035105} (\bibinfo{year}{2011}).

\bibitem{Eftekhari2012}
\bibinfo{author}{Eftekhari, F.} \& \bibinfo{author}{Davis, T.~J.}
\newblock \bibinfo{title}{Strong chiral optical response from planar arrays of
  subwavelength metallic structures supporting surface plasmon resonances}.
\newblock \emph{\bibinfo{journal}{Phys. Rev. B}} \textbf{\bibinfo{volume}{86}},
  \bibinfo{pages}{075428} (\bibinfo{year}{2012}).

\bibitem{Narushima2014}
\bibinfo{author}{Narushima, T.}, \bibinfo{author}{Hashiyada, S.} \&
  \bibinfo{author}{Okamoto, H.}
\newblock \bibinfo{title}{Nanoscopic study on developing optical activity with
  increasing chirality for two-dimensional metal nanostructures}.
\newblock \emph{\bibinfo{journal}{ACS Photonics}} \textbf{\bibinfo{volume}{1}},
  \bibinfo{pages}{732--738} (\bibinfo{year}{2014}).

\bibitem{Alali2014}
\bibinfo{author}{Alali, F.}, \bibinfo{author}{Kim, Y.~H.},
  \bibinfo{author}{Baev, A.} \& \bibinfo{author}{Furlani, E.~P.}
\newblock \bibinfo{title}{Plasmon-enhanced metasurfaces for controlling optical
  polarization}.
\newblock \emph{\bibinfo{journal}{ACS Photonics}} \textbf{\bibinfo{volume}{1}},
  \bibinfo{pages}{507--515} (\bibinfo{year}{2014}).

\bibitem{Plum2008}
\bibinfo{author}{Plum, E.}, \bibinfo{author}{Fedotov, V.~A.} \&
  \bibinfo{author}{Zheludev, N.~I.}
\newblock \bibinfo{title}{Optical activity in extrinsically chiral
  metamaterial}.
\newblock \emph{\bibinfo{journal}{Appl. Phys. Lett.}}
  \textbf{\bibinfo{volume}{93}}, \bibinfo{pages}{--} (\bibinfo{year}{2008}).

\bibitem{Plum2009}
\bibinfo{author}{Plum, E.} \emph{et~al.}
\newblock \bibinfo{title}{Metamaterials: Optical activity without chirality}.
\newblock \emph{\bibinfo{journal}{Phys. Rev. Lett.}}
  \textbf{\bibinfo{volume}{102}}, \bibinfo{pages}{113902}
  (\bibinfo{year}{2009}).

\bibitem{Sersic2011}
\bibinfo{author}{Sersic, I.}, \bibinfo{author}{Tuambilangana, C.},
  \bibinfo{author}{Kampfrath, T.} \& \bibinfo{author}{Koenderink, A.~F.}
\newblock \bibinfo{title}{Magnetoelectric point scattering theory for
  metamaterial scatterers}.
\newblock \emph{\bibinfo{journal}{Phys. Rev. B}} \textbf{\bibinfo{volume}{83}},
  \bibinfo{pages}{245102} (\bibinfo{year}{2011}).

\bibitem{DeLeon2015}
\bibinfo{author}{De~Leon, I.} \emph{et~al.}
\newblock \bibinfo{title}{Strong, spectrally-tunable chirality in diffractive
  metasurfaces}.
\newblock \emph{\bibinfo{journal}{Scientific Reports}}
  \textbf{\bibinfo{volume}{5}}, \bibinfo{pages}{13034} (\bibinfo{year}{2015}).

\bibitem{Yeom2013}
\bibinfo{author}{Yeom, B.} \emph{et~al.}
\newblock \bibinfo{title}{Chiral plasmonic nanostructures on achiral
  nanopillars}.
\newblock \emph{\bibinfo{journal}{Nano Lett.}} \textbf{\bibinfo{volume}{13}},
  \bibinfo{pages}{5277--5283} (\bibinfo{year}{2013}).

\bibitem{Hentschel2013}
\bibinfo{author}{Hentschel, M.}, \bibinfo{author}{Weiss, T.},
  \bibinfo{author}{Bagheri, S.} \& \bibinfo{author}{Giessen, H.}
\newblock \bibinfo{title}{Babinet to the half: Coupling of solid and inverse
  plasmonic structures}.
\newblock \emph{\bibinfo{journal}{Nano Lett.}} \textbf{\bibinfo{volume}{13}},
  \bibinfo{pages}{4428--4433} (\bibinfo{year}{2013}).

\bibitem{Evlyukhin2010}
\bibinfo{author}{Evlyukhin, A.~B.}, \bibinfo{author}{Reinhardt, C.},
  \bibinfo{author}{Seidel, A.}, \bibinfo{author}{Luk'yanchuk, B.~S.} \&
  \bibinfo{author}{Chichkov, B.~N.}
\newblock \bibinfo{title}{Optical response features of si-nanoparticle arrays}.
\newblock \emph{\bibinfo{journal}{Phys. Rev. B}} \textbf{\bibinfo{volume}{82}},
  \bibinfo{pages}{045404} (\bibinfo{year}{2010}).

\bibitem{Garcia2011}
\bibinfo{author}{Garc\'{i}a-Etxarri, A.} \emph{et~al.}
\newblock \bibinfo{title}{Strong magnetic response of submicron silicon
  particles in the infrared}.
\newblock \emph{\bibinfo{journal}{Opt. Express}} \textbf{\bibinfo{volume}{19}},
  \bibinfo{pages}{4815--4826} (\bibinfo{year}{2011}).

\bibitem{Kuznetsov2012}
\bibinfo{author}{Kuznetsov, A.~I.}, \bibinfo{author}{Miroshnichenko, A.~E.},
  \bibinfo{author}{Fu, Y.~H.}, \bibinfo{author}{Zhang, J.} \&
  \bibinfo{author}{Luk'yanchuk, B.}
\newblock \bibinfo{title}{Magnetic light}.
\newblock \emph{\bibinfo{journal}{Sci. Rep.}} \textbf{\bibinfo{volume}{2}}
  (\bibinfo{year}{2012}).

\bibitem{vandeGroep2013}
\bibinfo{author}{van~de Groep, J.} \& \bibinfo{author}{Polman, A.}
\newblock \bibinfo{title}{Designing dielectric resonators on substrates:
  Combining magnetic and electric resonances}.
\newblock \emph{\bibinfo{journal}{Opt. Express}} \textbf{\bibinfo{volume}{21}},
  \bibinfo{pages}{26285--26302} (\bibinfo{year}{2013}).

\bibitem{Wozniak2015}
\bibinfo{author}{Wozniak, P.}, \bibinfo{author}{Banzer, P.} \&
  \bibinfo{author}{Leuchs, G.}
\newblock \bibinfo{title}{Selective switching of individual multipole
  resonances in single dielectric nanoparticles}.
\newblock \emph{\bibinfo{journal}{Laser \& Photon. Rev.}}
  \textbf{\bibinfo{volume}{9}}, \bibinfo{pages}{231--240}
  (\bibinfo{year}{2015}).

\bibitem{Zhao2007}
\bibinfo{author}{Zhao, Y.}, \bibinfo{author}{Edgar, J.~S.},
  \bibinfo{author}{Jeffries, G. D.~M.}, \bibinfo{author}{McGloin, D.} \&
  \bibinfo{author}{Chiu, D.~T.}
\newblock \bibinfo{title}{Spin-to-orbital angular momentum conversion in a
  strongly focused optical beam}.
\newblock \emph{\bibinfo{journal}{Phys. Rev. Lett.}}
  \textbf{\bibinfo{volume}{99}}, \bibinfo{pages}{073901}
  (\bibinfo{year}{2007}).

\bibitem{Albella2013}
\bibinfo{author}{Albella, P.} \emph{et~al.}
\newblock \bibinfo{title}{Low-loss electric and magnetic field-enhanced
  spectroscopy with subwavelength silicon dimers}.
\newblock \emph{\bibinfo{journal}{J. Phys. Chem. C}}
  \textbf{\bibinfo{volume}{117}}, \bibinfo{pages}{13573--13584}
  (\bibinfo{year}{2013}).

\bibitem{Bakker2015}
\bibinfo{author}{Bakker, R.~M.} \emph{et~al.}
\newblock \bibinfo{title}{Magnetic and electric hotspots with silicon
  nanodimers}.
\newblock \emph{\bibinfo{journal}{Nano Lett.}} \textbf{\bibinfo{volume}{15}},
  \bibinfo{pages}{2137--2142} (\bibinfo{year}{2015}).

\bibitem{Bartenwerfer2011}
\bibinfo{author}{Bartenwerfer, M.} \emph{et~al.}
\newblock \bibinfo{title}{Towards automated afm-based nanomanipulation in a
  combined nanorobotic afm/hrsem/fib system}.
\newblock In \emph{\bibinfo{booktitle}{Mechatronics and Automation (ICMA), 2011
  International Conference on}}, \bibinfo{pages}{171--176}
  (\bibinfo{year}{2011}).

\bibitem{Mick2014}
\bibinfo{author}{Mick, U.}, \bibinfo{author}{Banzer, P.},
  \bibinfo{author}{Christiansen, S.} \& \bibinfo{author}{Leuchs, G.}
\newblock \bibinfo{title}{Afm-based pick-and-place handling of individual
  nanoparticles inside an sem for the fabrication of plasmonic nano-patterns}.
\newblock In \emph{\bibinfo{booktitle}{CLEO: 2014}}, \bibinfo{pages}{STu1H.1}
  (\bibinfo{publisher}{Optical Society of America}, \bibinfo{year}{2014}).

\bibitem{Banzer2010a}
\bibinfo{author}{Banzer, P.}, \bibinfo{author}{Peschel, U.},
  \bibinfo{author}{Quabis, S.} \& \bibinfo{author}{Leuchs, G.}
\newblock \bibinfo{title}{On the experimental investigation of the electric and
  magnetic response of a single nano-structure}.
\newblock \emph{\bibinfo{journal}{Opt. Express}} \textbf{\bibinfo{volume}{18}},
  \bibinfo{pages}{10905--10923} (\bibinfo{year}{2010}).

\bibitem{Yang2014}
\bibinfo{author}{Yang, A.} \emph{et~al.}
\newblock \bibinfo{title}{Hetero-oligomer nanoparticle arrays for
  plasmon-enhanced hydrogen sensing}.
\newblock \emph{\bibinfo{journal}{ACS Nano}} \textbf{\bibinfo{volume}{8}},
  \bibinfo{pages}{7639--7647} (\bibinfo{year}{2014}).

\bibitem{Tidwell1992}
\bibinfo{author}{Tidwell, S.~C.}, \bibinfo{author}{Ford, D.~H.} \&
  \bibinfo{author}{Kimura, W.~D.}
\newblock \bibinfo{title}{Transporting and focusing radially polarized laser
  beams}.
\newblock \emph{\bibinfo{journal}{Optical Engineering}}
  \textbf{\bibinfo{volume}{31}}, \bibinfo{pages}{1527--1531}
  (\bibinfo{year}{1992}).

\bibitem{ZhanBook2014}
\bibinfo{author}{Zhan, Q.}
\newblock \emph{\bibinfo{title}{Vectorial Optical Fields - Fundamentals and
  Applications}} (\bibinfo{publisher}{World Sceintific}, \bibinfo{year}{2014}).

\end{thebibliography}

\section*{Acknowledgments}
P. Banzer acknowledges financial support provided by the Alexander von Humboldt Foundation and the Canada Excellence Research Chair (CERC) in Quantum Nonlinear Optics. U. Mick acknowledges financial support by the DFG via the Research Training Group GRK1896. The authors acknowledge fruitful discussions with S. Orlov.

\section*{Author contributions}
P.B. conceived the idea; P.B. designed the structure and performed the numerical simulations; U.M. fabricated the sample; P.W. performed the experiment; P.B. and I.D.L. analyzed the data; R.W.B. and P.B. supervised all aspects of the project. All authors contributed to the preparation of the manuscript.

\section*{Competing financial interests}
The authors declare no competing financial interests.

\section*{Materials and Correspondence}
Correspondence to: P. Banzer, email: peter.banzer@mpl.mpg.de.
\pagebreak

\centerline{\textbf{SUPPLEMENTARY INFORMATION}}

\section*{Additional Experimental Results}
\begin{figure}
\includegraphics[width=0.85\textwidth]{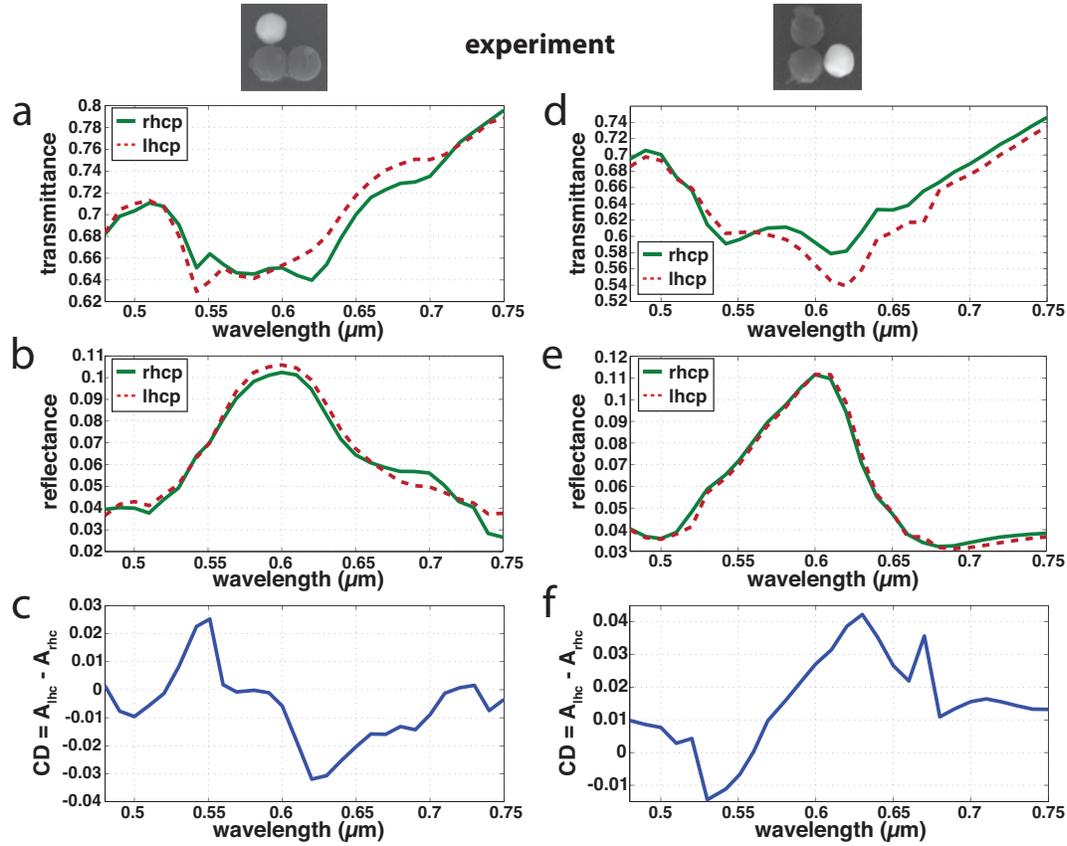}
\caption{\label{fig:FigS1}\textbf{Comparison of the optical response of two heterogeneous nanosphere trimers with opposite handedness.} (a)-(c) Experimental transmittance, reflectance and CD spectra for the trimer discussed in the main text. (d)-(f) Equivalent spectra for a trimer with opposite handedness. Deviations between corresponding spectra are a result of the slightly different particle sizes and arrangements.}
\end{figure}

To further prove the proposed concept of heterometerial selection for the creation of optically chiral nanostructures, we performed additional experiments on a trimer with mirrored geometry.  This trimer exhibits opposite handedness in comparison to the one discussed in the main text. The experiments were performed as described in the Methods section (see also Ref. \citen{Banzer2010a} for more details about the setup). In Fig. \ref{fig:FigS1}, the corresponding datasets are plotted (d - f) in comparison to the data for the trimer already discussed in the main text (a - c).

\end{document}